\documentclass{article}
\usepackage{graphicx} 
\usepackage{rotating}
\usepackage{tabularx}
\usepackage{booktabs}
\usepackage{array}
\usepackage{caption}
\usepackage{xcolor}
\usepackage{hyperref}
\usepackage[nobiblatex]{xurl}
\usepackage[square,numbers]{natbib}

\newcolumntype{L}[1]{>{\raggedright\arraybackslash}p{#1}} 
\newcolumntype{Y}{>{\raggedright\arraybackslash}X}        

\title{Foundations for Discovery: A Coordinated Fleet Approach to NASA Astrophysics}
\author{
\parbox{\textwidth}{
\centering
Regina Caputo, Francesca M. Civano, Knicole D. Col\'on, Brian Humensky, David T. Leisawitz, Avi M. Mandell, Conor A. Nixon, Georgia A. de Nolfo, Jeremy S. Perkins, Elisa V. Quintana, Judith L. Racusin, Joshua E. Schlieder, Albert Y. Shih, Amy A. Simon, Jacob Slutsky, Tonia M. Venters, Jennifer J. Wiseman, Allison A. Youngblood (NASA GSFC)
}
}

\date{September 2025}

\begin{document}

\maketitle

\tableofcontents

\newpage

\section{Executive Summary}

This white paper presents an analysis of Astro2020 science priorities and NASA's future astrophysics mission architecture, advocating for a coordinated fleet of \$1--2B missions, smaller than typical Flagship observatories, but strategically designed to complement them, i.e. a ``Next Generation Great Observatories" program. 
The study addresses opportunities in current mission planning, design, and implementation and proposes a strategic approach to maximize scientific return on investment while strengthening partnerships across NASA divisions, other government organizations, universities, and industry.

\subsection*{Key Findings}
The science priorities established by the Astro2020 Decadal Survey 
require capabilities that span multiple observational approaches and 
wavelength regimes. Our analysis demonstrates that most decadal 
science questions can be addressed through a coordinated fleet of 
$\sim$ \$1--2B  missions with a broad suite of scientific and 
technical capabilities and a significant guest investigator component, 
while operating concurrently with Flagship missions, to enable 
comprehensive, coordinated observations across the electromagnetic spectrum. 
Individually, these missions would deliver breakthrough, community serving science at lower cost than a Flagship. 
We thus introduce the concept of a~``Flaglet"~class mission. 
As an ensemble, a fleet of Flaglets could deliver science at Flagship scale. 

The key takeaways of a fleet of flaglet approach: 

\noindent \textbf{Coordination}: Develop a strategic program of Flaglet-scale missions that build exponentially on each other's capabilities while complementing Flagship missions, creating a diverse observational portfolio. 

\noindent \textbf{Cross-Divisional Synergies}: Leverage instrument 
design and operational strategies that are tuned for astrophysics but 
benefit planetary and heliophysics, maximizing utility across NASA science divisions and enhancing the scientific return of investments.

\noindent \textbf{Partnership Enhancement}: Strengthen collaborations with universities, industry, and other government agencies through coordinated mission planning and technological development that supports mission objectives at all scales. 
This can lead to huge cost savings throughout the mission life cycle. 

\subsection*{Achieving Transformational Science}
The proposed architecture prioritizes achieving the maximal number of decadal science goals and encompasses instrument and mission capabilities, spectral coverage, observatory response times, and technology readiness. 
This includes utilizing industry partnership models, common spacecraft bus and ground system strategies, ride-share opportunities, and instrument competitions to drive innovation while supporting the portfolio of mission goals. As a critical next step, we advocate for a large, community-led mission study under NASA's Astrophysics Strategic Technology \& Research Accelerator (ASTRA) initiative to  define the Flaglet fleet architecture, with community engagement activities anticipated to begin as early as mid-2026.
The approach promises enhanced program efficiency and increased scientific return through mission synergies while maintaining strong alignment with Astro2020 priorities.

\section{Introduction}

\textbf{NASA’s Great Observatories program} is arguably the most successful paradigm for comprehensive study of the universe in human history.  Conceived in the 1970’s long before the first Great Observatory, the Hubble Space Telescope, was launched, the idea was to establish a suite of major observatories in space, thus unhindered by atmospheric interference, to cover multiple parts of the electromagnetic spectrum and study a broad swath of astronomical phenomena.  The National Research Council’s ``Decadal'' Survey recommendations for the 1980’s set the stage for advocacy by multiple NASA HQ leaders, with the rationale that future large space observatories should complement rather than replace one another, and should  be thought of as one coherent panchromatic platform to study everything from the solar system to cosmology.

One of the NASA documents promoting the concept concluded:\\\textit{} 
\begin{quote}
“In the Space Station era, the family of permanent observatories in space will open the way to new, comprehensive studies of key remaining problems in astrophysics, helping us understand:
The birth of the Universe, its large-scale structure, and the formation of galaxies and clusters of galaxies;  The fundamental laws of physics governing cosmic processes and events;  The origin and evolution of stars, planetary systems, life and intelligence.
\textbf{If we succeed, we will leave a legacy to rank us with the great civilizations of the past.”}~\cite{harwit1986}
\end{quote}

\begin{figure}[ht]
    \centering
    \includegraphics[width=0.95\linewidth]{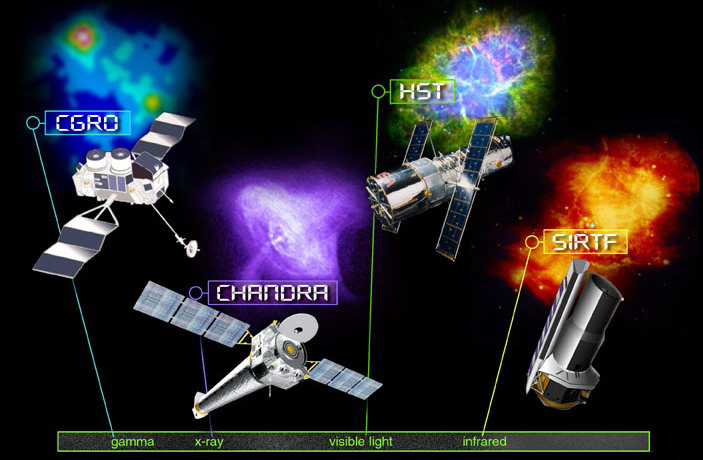}
    \caption{NASA’s original Great Observatory Suite, placed in its multiwavelength context via the incredibly diverse views provided of the Crab Nebula.}
    \label{fig:fleet}
\end{figure}

\noindent With the launches of the Hubble Space Telescope (HST, 1990), the Compton Gamma-Ray Observatory (CGRO, 1991), the Chandra X-ray Observatory (1999), and the Spitzer Space Telescope (a.k.a SIRTF, 2003), (\textbf{Figure~\ref{fig:fleet}}) profound discoveries about the universe previously unimaginable have been achieved, revealing phenomena ranging from solar system dynamics to exoplanets to cosmic explosions in the early universe.   
Four decades of powerful discoveries, often relying on synergistic study across multiple wavelength bands covered by the Great Observatory suite, have literally changed textbooks and humanity’s view of the cosmos and our place within it.

Over the decades, this phenomenally successful paradigm of a suite of major permanent platforms together accessing multiple wavelength bands has not changed, but the components within it have:  the CGRO mission ended, Spitzer was eventually replaced by the more powerful James Webb Space Telescope (JWST), and the individual science instruments on Hubble have been replaced and upgraded multiple times with astronaut servicing missions. The Roman Space Telescope (RST), a new Flagship observatory, will soon add to the palette with revolutionary wide-field surveys and breakthrough technology demonstrations. And on the more distant horizon, the Habitable Worlds Observatory (HWO) would succeed Hubble’s ultraviolet-visible-near-IR coverage with a larger more powerful platform.  

Other advances over the decades are now influencing how we envision the future of space-based astrophysics.   
The professional astronomical community has learned that strong scientific advancement focused on specific observational goals is possible through a litany of small and medium class astronomy missions.   
We have also been steadily advancing technology, such that powerful astronomical detectors and observatory platforms can now be developed at a lower cost point than previously possible through the use of suborbital platforms~\cite{suborbitals}. 
The most recent National Academy Decadal Survey~\cite{NAP26141} pointed out that much of their recommended science program requires a panchromatic fleet and a broad program:\\

\begin{quote}
“... missions of all scales, national and international, designed to view the universe 
in a multiplicity of complementary ways are now essential to progress in modern astrophysics.” ~\cite{NAP26141}   

\end{quote}

Thus looking toward the future, and continuing the concept of the Great Observatories paradigm, advances in technology along with expanding commercial capabilities could enable a multi-wavelength suite of ground-breaking science mission opportunities at a lower cost than a historical large Flagship mission, as evidenced by Probe-class mission white papers submitted to the Decadal Survey panel.   
This report presents a summary of the range of high priority science goals that a suite of such lower-cost missions could address.\\


\subsection{Looking Ahead: A New Future Great Observatories Pathway}
 
\begin{figure}[htb!]
    \centering
    \includegraphics[width=0.95\linewidth]{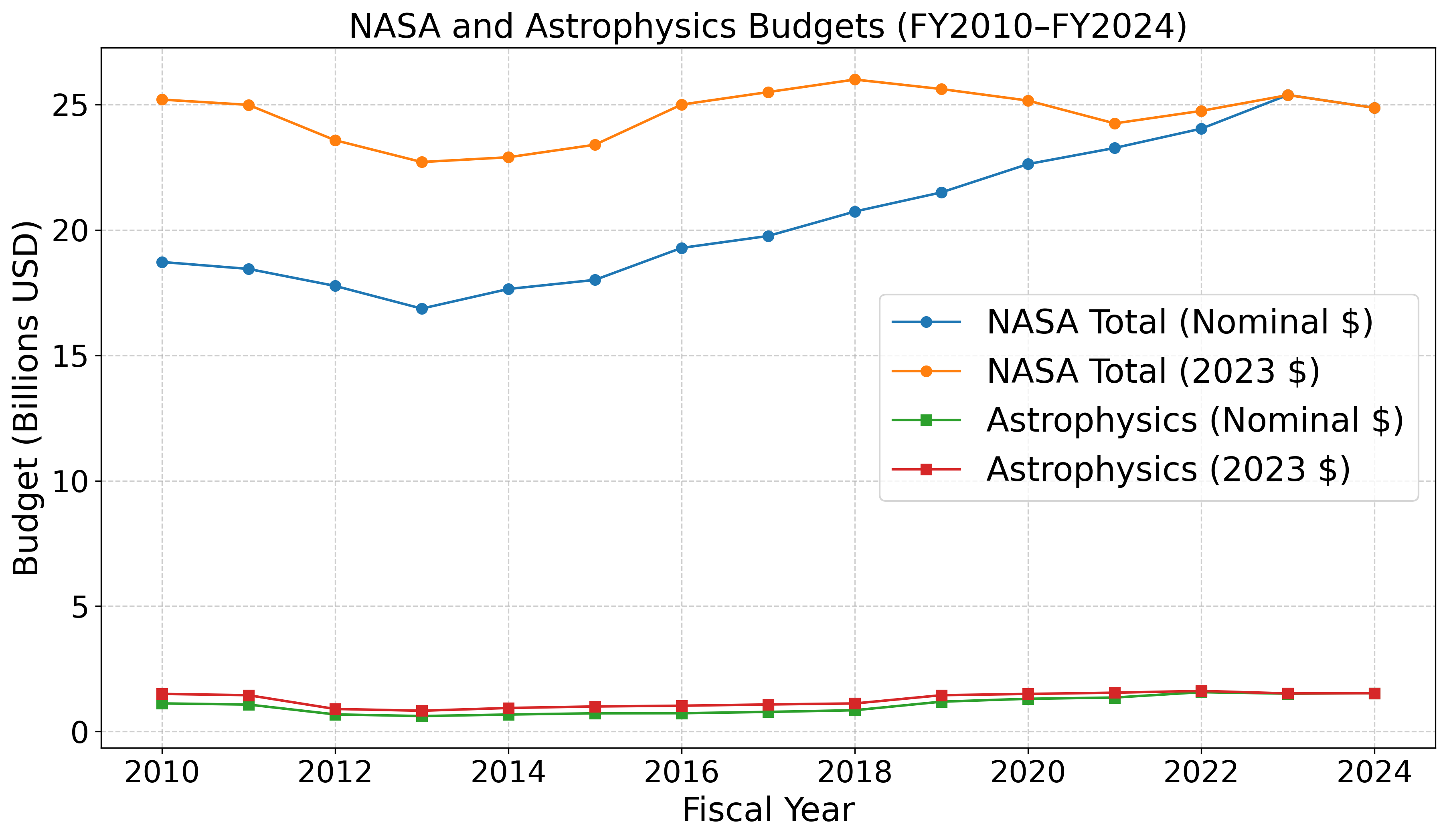}
    \caption{NASA overall and Astrophysics budgets for fiscal years 2010 to 2024, including the inflation adjusted values (scaled to 2023 dollars). Overall, the NASA budget has been flat for over 15 years. Nominal numbers for NASA total and Astrophysics come from NASA budget documents collected by The Planetary Society~\cite{planetarysociety} }
    \label{fig:budget}
\end{figure}

Advances in technology, current fiscal constraints (\textbf{Figure~\ref{fig:budget}}), and a wide variety of meritorious ideas suggest that the most expedient strategy for achieving the ground-breaking science outlined in the Astro2020 Decadal Survey~\cite{NAP26141} may be a comprehensive, planned, panchromatic fleet of \$1-2B missions, smaller than a traditional Flagship (hence, Flaglet), yet powerful in capability. 
Not only would missions of this scale change the state of each of their fields at a cadence commensurate with the span of a person's career, but a planned program with complementary capabilities would exponentially build on each other to achieve maximum return on investment, while also bolstering partnerships with universities, industry and other government agencies, all while fundamentally complementing the science objectives of current and future Flagships.  

The value of this scale of mission is already outlined in the Astro2020 Decadal Survey which calls for ``a new line of Probe missions to be competed in broad areas identified as important to accomplish the survey’s scientific goals". 
Beyond the recent Astrophysics Probe Explorer Announcement of Opportunity (APEX AO), no decision has been made (or recommended) on the structure and scientific objectives of this mission class going forward. 
Furthermore, Astrophysics has had a long history of providing critical observations that connect back to our local solar system and heliosphere. 
When defining a future fleet of observatories, it is important to consider instrument design and operational trades that would benefit planetary and solar/heliospheric observations. 

\begin{figure}[ht]
    \centering
    \includegraphics[width=0.95\linewidth]{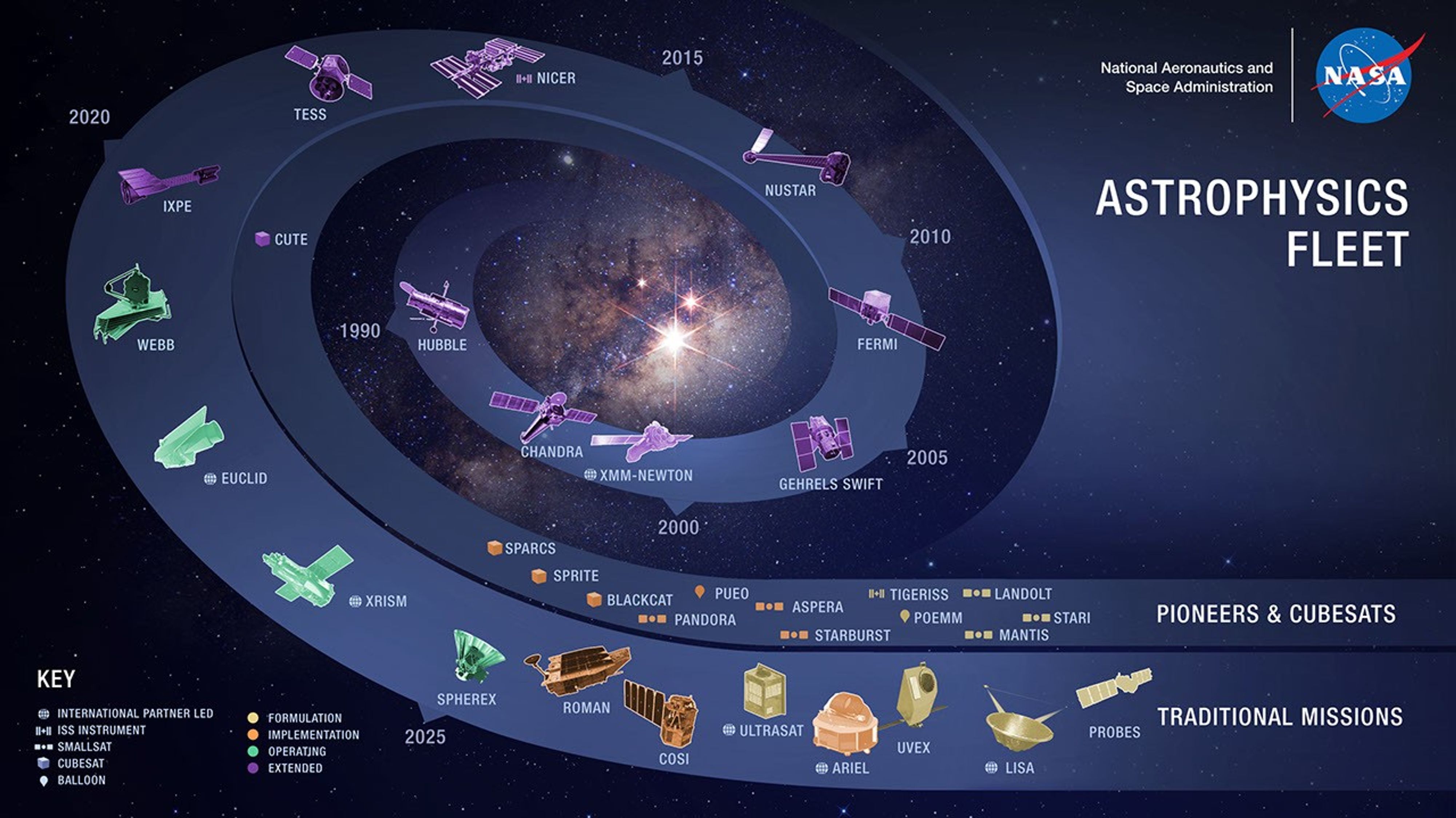}
    \caption{The current fleet of NASA Astrophysics missions~\cite{fleet} showing launch, mission phase, and scale across different wavelength regimes. The complementary nature of these missions creates a robust observational framework that we aim to build upon with the strategic approach detailed in this document.}
    \label{fig:snail}
\end{figure}

The current fleet of Astrophysics missions, shown in \textbf{Figure~\ref{fig:snail}}, demonstrates operational observatories across different wavelength regimes. 
It exemplifies the existing coverage and capabilities of NASA's space-based astrophysics program, highlighting both the strengths of current assets and potential gaps that could be addressed by Flaglet missions, particularly when looking a decade in the future. 
The complementary nature of these missions creates a robust observational framework that the strategic approach described here seeks to build upon.

\section{Methodology}\label{sec:methodology}

The Astro2020 Decadal Survey~\cite{NAP26141}, was a massive community effort to identify and prioritize the science goals of the Astrophysics community over the subsequent decade. 
The goal of our team was to investigate a path to maximally address these science goals while building an efficient and cost-effective program. 
Our team systematically matched the capabilities required to address these science goals with the capabilities provided by Probe scale mission concept white papers submitted to Astro2020.
We are informed by the legacy of the four Great Observatories (Hubble, Chandra, Compton, and Spitzer). 
Our methodology centers on identifying decadal science gaps, assessing the potential of each proposed Probe mission to fill the gaps, and determining whether a small number of Flaglet missions at comparable scale, represented by their measurement capabilities, could collectively accomplish a large fraction of the science.

\subsection{Science Gap Analysis}

We conducted a detailed review of the \textit{Astro2020 Decadal Survey} to extract high-priority science goals that are currently underserved by the strategic Flagship missions JWST, RST, and HWO. 
Each broad decadal survey theme was treated separately: \textit{Worlds \& Suns in Context} (WSC), \textit{Cosmic Ecosystems} (CE), and \textit{New Messengers \& New Physics} (NMNP).
Within each theme, each question prioritized by the six decadal scientific panels was examined. 
In an initial review, it quickly became apparent that accomplishing these questions requires multi-wavelength observations across the electromagnetic spectrum, and in several cases multi-messenger particle and gravitational wave observations. 
These questions were cross-referenced with missions currently planned and in development (e.g., RST, HWO) to isolate unmet needs and identify gaps.

\subsection{Multi-Wavelength/Messenger Requirement Mapping}

For each identified science theme, we mapped the necessary observational capabilities to understand where the Probe-scale mission concepts submitted to the decadal survey could close existing coverage gaps. 
This step involved an analysis of the measurement capabilities of the observatories submitted as part of the Probe-concept study reports.
We also compiled a set of illustrative science questions, from the appendices of the decadal survey (one from each of the Appendices B-G, with representation from each topical group), in each science theme and compiled the capabilities required to address each question. 
Furthermore, we reached out to the \href{https://axis.umd.edu/}{AXIS} and \href{https://prima.ipac.caltech.edu/}{PRIMA} teams -- two missions in Phase A as part of the APEX AO -- to cross reference their capabilities and validate our methodology. 

\subsection{The Utility of \$1-2B Missions}
\label{sec:utility}
We cross correlated which current and planned strategic Flagships and Probe-scale mission concepts could address each observational capability. 
Each Flagship and concept was assessed on its ability to address the capabilities needed to achieve decadal science questions, using the categories ``none'', ``somewhat'', and ``mostly/entirely''. We then determined how many more questions could be addressed beyond the current and planned strategic missions to illustrate the utility of this class within the NASA fleet.  
\textbf{Our aim was not to rank individual missions but rather to determine whether a small number of them with representative capabilities could fill a large fraction of the science gaps.}
Our analysis demonstrated how many decadal survey questions in each theme could be fully or partially addressed with additional Flaglet-class missions.

\subsection{Architectural and Operational Trade Studies}

It is beyond the scope of this work to investigate potential architectural and operational trade studies such as: instrument and mission capabilities, observatory response times, technology readiness and modularity (such as configurable spacecraft buses and common ground systems), and potential synergies with ongoing NASA programs and the upcoming HWO. 
Our goals are to identify a path to enhancing program efficiency, scientific return, and cross-disciplinary utility. 
Further studies should include evaluating industry partnership models for spacecraft development, common bus or platform strategies, and additional opportunities for cross-divisional coordination with the planetary and heliophysics communities.
This could be accomplished via a large mission concept study in preparation for Astro2030.

\subsection{Findings}
\label{sec:findings}

Our analysis of the decadal survey science priorities and current mission concepts reveals several key insights that could inform the astrophysics program.
In particular, we found that most science questions require more capabilities than one single observatory can provide. 
The most efficient way to address these priorities is developing a fleet of more modest in scope (\$1-2B  Flaglets) with multi-wavelength/messenger capabilities that are launched close-in-time so they have overlapping observations.
This is consistent with studies on the scientific productivity of different mission scales~\cite{flaglets}.  
Our gap analysis demonstrated that while individual missions excel in specific wavelength regimes or observational techniques, the complex, multi-faceted questions prioritized by the community demand coordinated, multi-wavelength/messenger approaches.

\subsubsection{Fleet Architecture Advantages}
The architectural trade studies presented in detail in Section~\ref{sec:capabilities} consistently pointed toward a distributed approach rather than singular large observatories. 
Our analysis shows that a coordinated fleet of Flaglet-scale missions offers several strategic advantages:

\begin{itemize}
    \item \textbf{Enhanced scientific return}: Multiple observatories with overlapping observation windows can address a broader range of decadal questions simultaneously
    \item \textbf{Risk mitigation}: Distributed capabilities reduce single-point-failure risks
    \item \textbf{Programmatic flexibility}: Staggered development timelines allow for technology maturation and incorporation of lessons learned 
    \item \textbf{Community access}: Multiple platforms provide more observing opportunities across the broad astrophysics community
\end{itemize}

\subsubsection{Multi-wavelength Synergy Requirements}
A critical finding from our mission capability assessment is that the highest-impact science emerges from coordinated observations across wavelength regimes. 
The science questions identified as top priorities by the decadal survey consistently require simultaneous or near-simultaneous observations in multiple bands, from gamma-ray through radio, that are technically and programmatically impossible to achieve within a single observatory.

\subsubsection{Implementation Pathway}
Based on these findings, \textbf{the science prioritized by the decadal survey should be driving the requirements and capabilities for missions}. 
The optimal path forward involves developing a fleet of specialized but complementary observatories with multi-wavelength/messenger capabilities that are launched close-in-time to ensure overlapping operational periods. 
This approach maximizes scientific return while maintaining programmatic feasibility and cost-effectiveness.
These findings, detailed further in Section~\ref{sec:science}, form the foundation for our strategic recommendations outlined in the following sections.

\subsection{Community Engagement and Dissemination}

To ensure broad community input and support, we have summarized our findings in this report and made it publicly available. 
We plan to present our results at relevant professional society meetings, including the American Astronomical Society (AAS), and may seek to organize special or splinter sessions in collaboration with NASA's PhysCOS, COR, and ExEP program offices. 
These forums will be used to solicit feedback, promote cross-sector collaboration, and refine the vision for a Flaglet observatory fleet.
We also suggest to further refine this analysis via a large mission study, as part of the recently announced Astrophysics Strategic Technology \& Research Accelerator (ASTRA) Initiative\footnote{\url{https://science.nasa.gov/astrophysics/programs/physics-of-the-cosmos/studies/astra-initiative/}}. 

\section{Capabilities to Maximize Science Return}
\label{sec:capabilities}

Our analysis focused on determining the fewest number of missions that would be needed to accomplish a given number of decadal survey science goals while striving to maximize the science return. 
We present our findings without naming specific mission concepts but rather by focusing on the measurement capabilities similar missions would bring if they were realized. 
The results of our analysis are shown in \textbf{Figures~\ref{fig:s-curves_all} \&~\ref{fig:s-curves_some}}. 
\textbf{Figure~\ref{fig:s-curves_some}} shows that, after JWST, RST, and HWO, three additional Flaglet class missions would fully or partially address at least 80\% of the outstanding questions in {\it Cosmic Ecosystems} and four would do so in {\it New Messengers New Physics}, with diminishing returns as additional Flaglets are added. After HWO, $\sim$80\% of the {\it Worlds and Suns in Context} science questions will already have been answered. We also found overlap in the set of missions that optimize science return from {\it Cosmic Ecosystems} and {\it New Messengers New Physics}; the capabilities of at most six of the Probe mission concepts would enable the accomplishment of at least 80\% of the science.
This is a remarkable finding since these missions were all developed well in advance of the decadal survey recommendations. 

In \textbf{Table~\ref{tab:capabilities}} we summarize a set of measurement capabilities across wavelengths and messengers that, if collectively available, would answer a significant fraction of the outstanding Astro2020 priority science questions. 
The capabilities described in each row of the table are realizable in a proposed mission in the \$1-2B class. While the mission concepts were not originally conceived this way, in some cases a single mission could offer capabilities described in more than a single table row, reducing the number of needed new missions to fewer than six if strategic design choices were made.

In summary, we find that \textbf{that a relatively small number (4) of Flaglet-class missions can answer a large fraction of the decadal-prioritized science questions.}
This set of missions would be the foundation for the Future Great Observatories.

\begin{table}[ht!]
\centering
\caption{Capabilities across different wavelength, energy, and messenger regimes for a panchromatic observatory fleet.}
\renewcommand{\arraystretch}{1.2}
\begin{tabularx}{\textwidth}{>{\raggedright\arraybackslash}p{1in} >{\raggedright\arraybackslash}p{1.2in} X}
\toprule
\textbf{Messenger} & \textbf{Wavelength / Energy} & \textbf{Key Capabilities} \\
\midrule
Neutrinos & PeV -- EeV & High-energy neutrino observations \\
Cosmic Rays & EeV & Ultra-high-energy cosmic ray observations \\
High Energy Gamma Rays & 100 MeV -- 100 GeV & Wide field-of-view all-sky monitoring; spectroscopy; imaging; polarimetry \\
Medium Energy Gamma Rays & 100 keV -- 100 MeV & Wide field-of-view all-sky monitoring; spectroscopy; imaging; polarimetry \\
Hard X-rays & 10 -- 100 keV & Wide field-of-view monitoring; rapid response; imaging and spectroscopy \\
Soft X-rays & 0.1 -- 10 keV & Spectral timing; Wide field-of-view monitoring; rapid response; imaging and spectroscopy; polarimetry \\
UV & 0.012 -- 0.4 $\mu$m & Wide field-of-view imaging; rapid response; spectroscopy and imaging;  \\
Optical & 0.4 -- 0.7 $\mu$m & Wide field-of-view imaging; rapid response; high-contrast imaging and spectroscopy; astrometry; polarization \\
NIR & 0.7 -- 2 $\mu$m & High-contrast imaging and spectroscopy; astrometry; polarization \\
Mid-IR & 2 -- 20 $\mu$m & Ultra-stable eclipse spectroscopy; high-contrast imaging \\
Far-IR & 20 -- 450 $\mu$m & Sub-arcsecond imaging; R$>$3000 integral field spectroscopy; 10 $\mu$Jy sensitivity \\
Sub-mm/ Microwave & 450 $\mu$m -- 3 cm & Background limited polarization imaging \\
\bottomrule
\end{tabularx}

\label{tab:capabilities}
\end{table}

\subsection{Role of New Technology and Commercial Capabilities}

Generally, the capabilities of the Probe-class missions submitted to the Decadal Survey are enabled by new technologies. 
Some proposed technologies have already reached or are approaching Technology Readiness Level (TRL) 6 for the missions in question, while others require further investment. 
This investment could come by expanding the Great Observatories Maturation Program (GOMAP) envisaged in the Decadal Survey, beyond HWO and other Flagships, and dedicate part of it to maturing technologies for \$1-2B missions whose capabilities appear in \textbf{Table~\ref{tab:capabilities}}. 
These investments should be ``pulled” by notional missions and sustained until the enabling technologies reach TRL 6.

A Future Great Observatories program, achieved through a coordinated fleet of Flaglets, will benefit substantially through cost savings and risk reduction traceable to industry-offered technologies. This includes spacecraft buses developed for non-NASA customers and launch capabilities that were not available during the era of the original Great Observatories. Modern commercial spacecraft buses, originally developed for telecommunications, Earth observation, and other commercial applications, now offer standardized platforms with sufficient power and data-handling capabilities to support sophisticated astrophysics instruments. 
These platforms have accumulated thousands of flight hours across multiple missions, providing a proven foundation that reduces both development time and technical risk.

Today, new launch vehicles with enhanced capabilities enable much larger and more massive payloads at greatly reduced cost. Heavy-lift vehicles like Falcon Heavy and New Glenn, as well as other emerging capabilities, offer payload capacities and fairing volumes that were previously available only through custom, expensive launch solutions. This enhanced lift capability enables larger apertures, more comprehensive instrument suites, and reduced mass constraints that historically drove complex engineering solutions. Leveraging ``economy of scale'' principles presents multiple strategic advantages.  Designing missions around generic, commercially-available spacecraft buses reduces the need for custom bus development, shortening development timelines and reducing costs. A coordinated Future Great Observatory program spanning the entire panchromatic spectrum could standardize interfaces, operations procedures, and ground systems, creating additional efficiencies through shared infrastructure and expertise.

The commercial communications industry further amplifies these benefits by providing robust, high-bandwidth data relay capabilities and standardized communication protocols for observatories in Earth orbits. NASA is studying and investing in these future capabilities for the NASA fleet via the Communications Services Project. 
This infrastructure reduces the need for dedicated ground stations and enables more flexible mission operations, including coordinated observations between multiple observatories. 

The commercial space sector has matured dramatically, creating unprecedented opportunities for NASA to leverage established, flight-proven technologies and infrastructure. These new commercial capabilities, combined with standardized spacecraft platforms, create an unprecedented opportunity to deploy a coordinated fleet of Flaglet-size missions that collectively provide comprehensive astrophysical coverage at lower cost than traditional development approaches.

\subsection{Capability Gaps}

Despite the broad range of mission concepts submitted as white papers to the Astro2020 decadal survey, several observational capabilities were absent from Probe-scale mission proposals. 
Missing capabilities include 21-cm line mapping missions designed to observe neutral hydrogen at redshifts $z \sim 6$--$12$ across tens of square degrees or more, which are essential for probing the epoch of reionization. 
The absence of submitted 21-cm interferometer missions, 
along with the lack of advanced radio interferometry mission concepts, could suggest either the expectation that these observations could be accomplished via ground based experiments or potential gaps in the white paper submissions. 
It is important to note that mission concepts addressing these capabilities may well exist within the broader astrophysics community but were not submitted to the decadal survey for various reasons and investigating these submission gaps was beyond the scope of this analysis.

\section{Science Achievable by a Fleet of Flaglet-scale Missions} 
\label{sec:science}

A fleet of large (\$1-2B) astrophysics missions spanning the electromagnetic spectrum represents a transformative approach to addressing the majority of Astro2020 decadal survey priorities through comprehensive multi-wavelength capabilities. The science questions this scale of observatories would simultaneously tackle span all the decadal survey panels. 

\begin{figure}[ht]
    \centering
    \includegraphics[width=\textwidth]{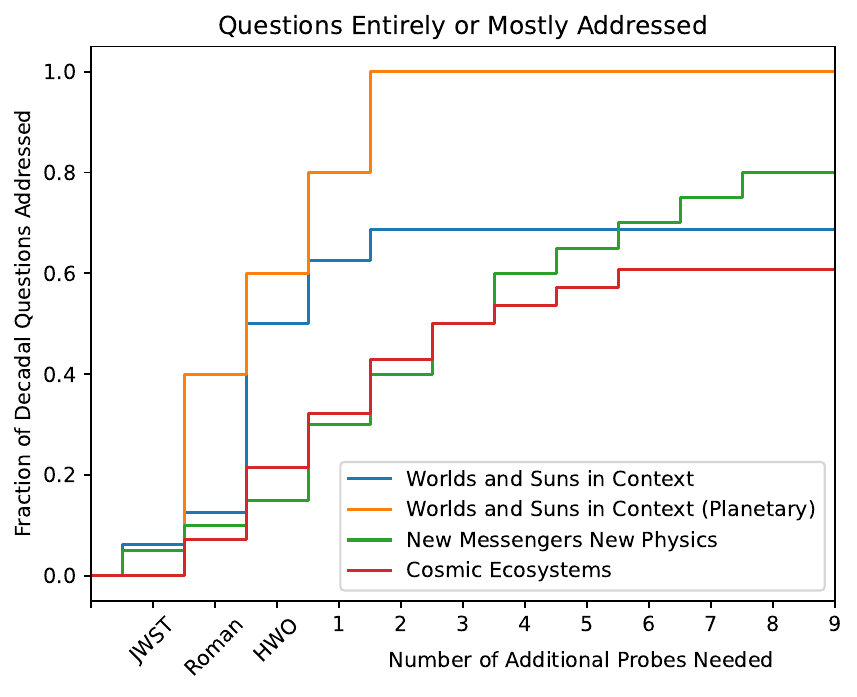}
    \caption{With the addition of $\sim$4-6 \$1-2B missions to current or planned (adopted) Flagship missions, we can entirely or mostly address a large proportion of Astro2020 decadal survey \cite{NAP26141} science panel questions. These criteria were derived from the analysis described in Section~\ref{sec:methodology} that studied the decadal science questions and required capabilities addressed by adopted missions and Probe-class mission concepts.}
    \label{fig:s-curves_all}
\end{figure}

\begin{figure}[ht]
    \centering
    \includegraphics[width=\textwidth]{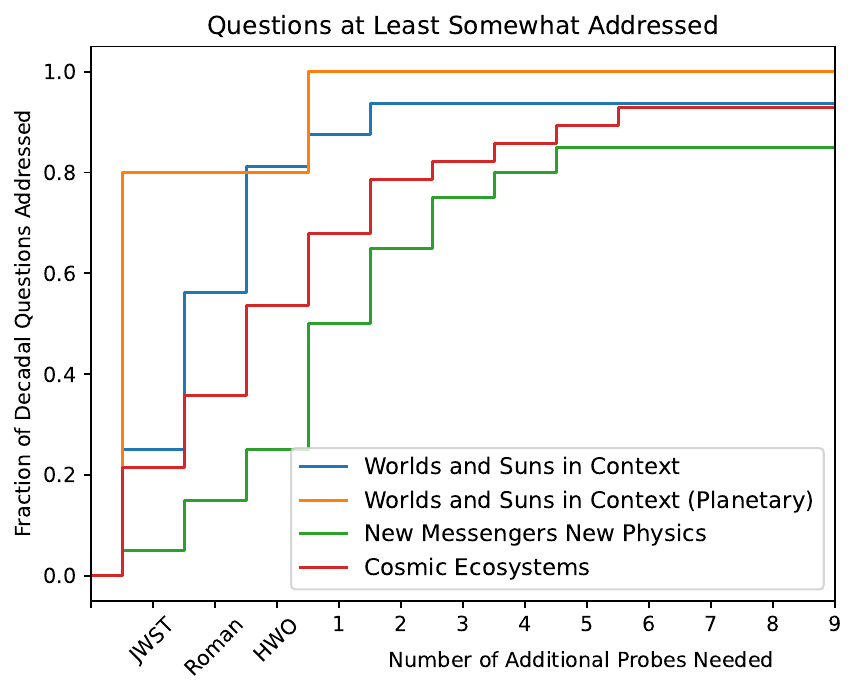}
    \caption{With the addition of $\sim$4 \$1-2B missions to current or planned (adopted) Flagship missions, we can at least somewhat address nearly all Astro2020 decadal survey \cite{NAP26141} science panel questions. These criteria were derived from the analysis described in Section~\ref{sec:methodology} that studied the decadal science questions and required capabilities addressed by adopted missions and Probe-class mission concepts.}
    \label{fig:s-curves_some}
\end{figure}

In our review of decadal science questions and Probe-concept capabilities, once we isolated the core capabilities, we identified the concepts within each science theme that addressed the largest number prioritized capabilities that are not covered by the strategic program of record (JWST, RST, and HWO). Independent of the details of the concept or capability, we assigned the first concept an index of 1, and repeated the above exercise on the remaining concepts and capabilities not addressed by the program of record or Probe-concept 1 (yielding Probe-concepts 2, 3, 4, etc.)  
We iteratively identified the minimum number of concepts to address the science questions in each section. 
\textbf{Figures \ref{fig:s-curves_all}} and \textbf{\ref{fig:s-curves_some}} depict the fraction of relevant decadal questions wholly or partially answered by the program of record plus additional Probe-class missions for each subject area, respectively.

\subsection{Astro2020 Science Questions}

To illustrate our methodology, we have selected one representative question from each decadal science panel (labeled as (`letter'-Q`number')) to demonstrate the efficacy of a suite of Flaglet-scale missions (Table \ref{tab:questions}).

\paragraph{Why do some compact objects eject material in nearly-light-speed jets, and what is that material made of? (B-Q3):} Many astrophysical systems create collimated beams of intense radiation and matter accelerated to close to the speed of light. These ``jets'' have been observed in several source classes, particularly those that contain or form compact objects such as supermassive or stellar-mass black holes, neutron stars, and perhaps even white dwarfs. 
Given the variety of conditions that give rise to them, relativistic jets serve as unparalleled laboratories for a wide range of energetic phenomena that are discernible across the electromagnetic spectrum, from low-frequency radio waves to the highest-energy gamma rays, and even through new messengers such as gravitational waves, high-energy neutrinos, and ultra-high-energy cosmic rays. However, despite the prevalence of relativistic jets, many facets of their energetic phenomena remain a mystery. 

\begin{figure}[ht]
    \centering
    \includegraphics[width=0.95\linewidth]{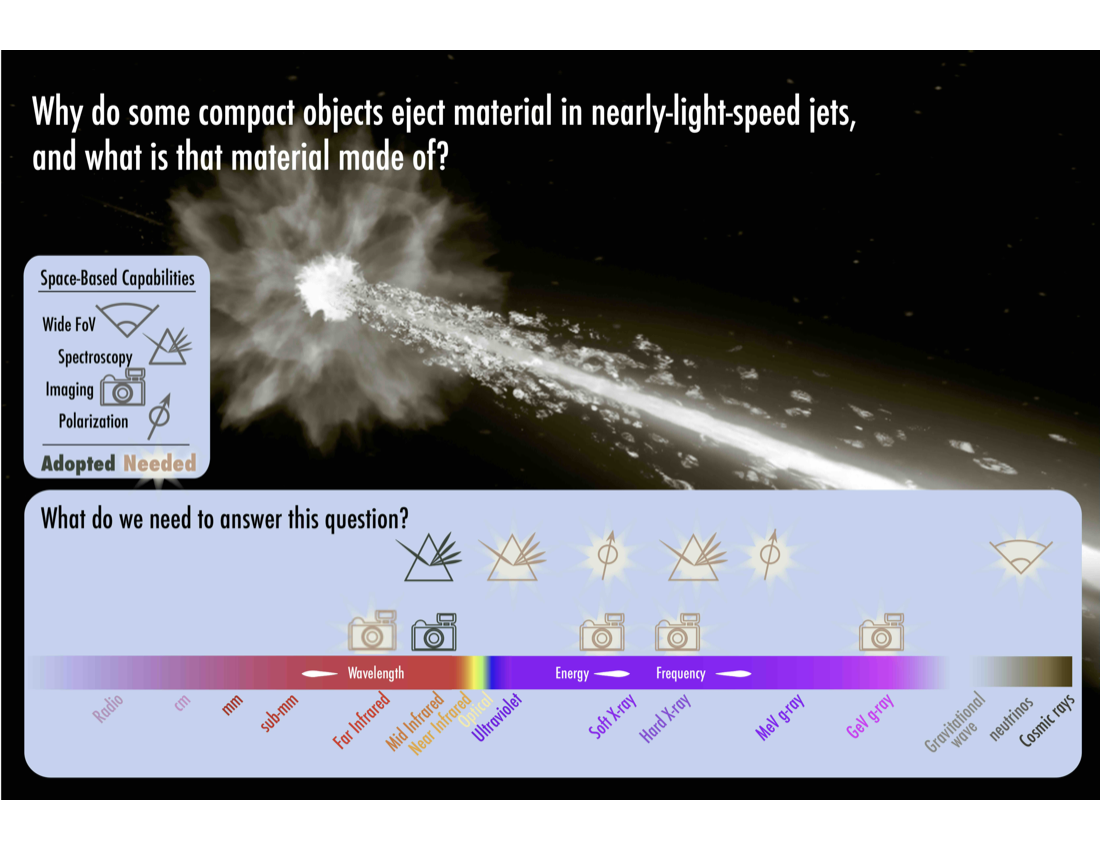}
    \caption{The Astro2020 Decadal science sub-panels outlined the most important scientific questions to be addressed by specific capabilities.  We outline Question B-Q3 from the \textit{New Messengers New Physics} theme and the panchromatic capabilities required to answer it. The optical and NIR capabilities are provided by JWST, RST, and HWO, but the high-energy and multimessenger capabilities can be addressed by a coordinated strategic fleet of Flaglet missions.}
    \label{fig:ex1_capabilities}
\end{figure}

A cohesive fleet of observatories with capabilities reaching across different wavelengths, measurement techniques, and messengers, such as the program being proposed here, would provide unparalleled opportunities to answer the many open questions about the nature of relativistic jets (illustrated in Figure~\ref{fig:ex1_capabilities}). 
High-resolution imaging in optical, infrared, and X-rays will probe jets in the acceleration and collimation regions that are complementary to those achievable in radio. At the same time, these observations will allow searches for re-collimation shocks in the ejection of new components that would be more closely linked with gamma-ray variability. 
X-ray and gamma-ray polarimetry will provide information about jet magnetic field configurations and orientations and distinguish between leptonic and hadronic radiative processes. 
Wide-field monitoring in soft X-rays through MeV gamma rays, as well as continued monitoring in the GeV band, will provide time-resolved measurements of jet characteristics crucial to understanding their evolution with time. Extending neutrino searches to PeV and EeV energies will be necessary to determine whether jets accelerate baryons to the very-high and ultra-high energies that produce these neutrinos. Large-exposure experiments that observe ultra-high-energy cosmic rays will provide the necessary statistics to finally pinpoint their sources and determine if they can be linked to relativistic jets.

\paragraph{What are the properties of dark matter and the dark sector? (C-Q2):}
The field of dark matter theory and detection has undergone a paradigm shift. In previous decades, the field focused primarily on two dark matter candidates, weakly interacting massive particles (WIMPs) and axions, motivated mainly by their ability to solve long-standing open questions within the Standard Model of particle physics. However, recent work has emphasized that dark matter may arise from a dark sector that is more analogous to the visible sector of familiar particles, with its own dynamics and forces leading to new terrestrial and astrophysical signatures. 
The non-detection of physics beyond the Standard Model at the Large Hadron Collider (LHC) — not finding signatures of supersymmetry — has served to further highlight the possibility that the dark sector need not be closely connected to well-recognized questions in particle physics. The breadth of new dark matter candidates and dark sector dynamics that have arisen from these recent explorations renews motivation and opportunities to detect astrophysical signatures of dark matter.

The search for dark matter signatures is wide-ranging and exploratory, but the next generation of radio telescopes for pulsar timing, large-aperture optical telescopes, high-resolution CMB polarization mapping, GeV telescopes, and TeV-scale Cherenkov telescopes are particularly important to make progress in this field. 
Several of these core capabilities that can lead to the next breakthroughs in dark matter science can be achieved by the synergistic Flaglet-size missions architecture described here.

\paragraph{How do supermassive black holes form, and how is their growth coupled to the evolution of their host galaxies? (D-Q3):} 
Observations during the last 20 years have shown that supermassive black holes reside at the center of nearly all galaxies. 
Tight correlations between host galaxy properties and supermassive black hole properties have shown that the evolution of the two are strongly connected and correlated. 
However, several major questions related to the formation of black holes in the early universe, their growth in this epoch, their role in re-ionization, and the processes driving those strong correlations (a.k.a feedback) are not yet fully understood. 

\begin{figure}[ht]
    \centering
    \includegraphics[width=0.95\linewidth]{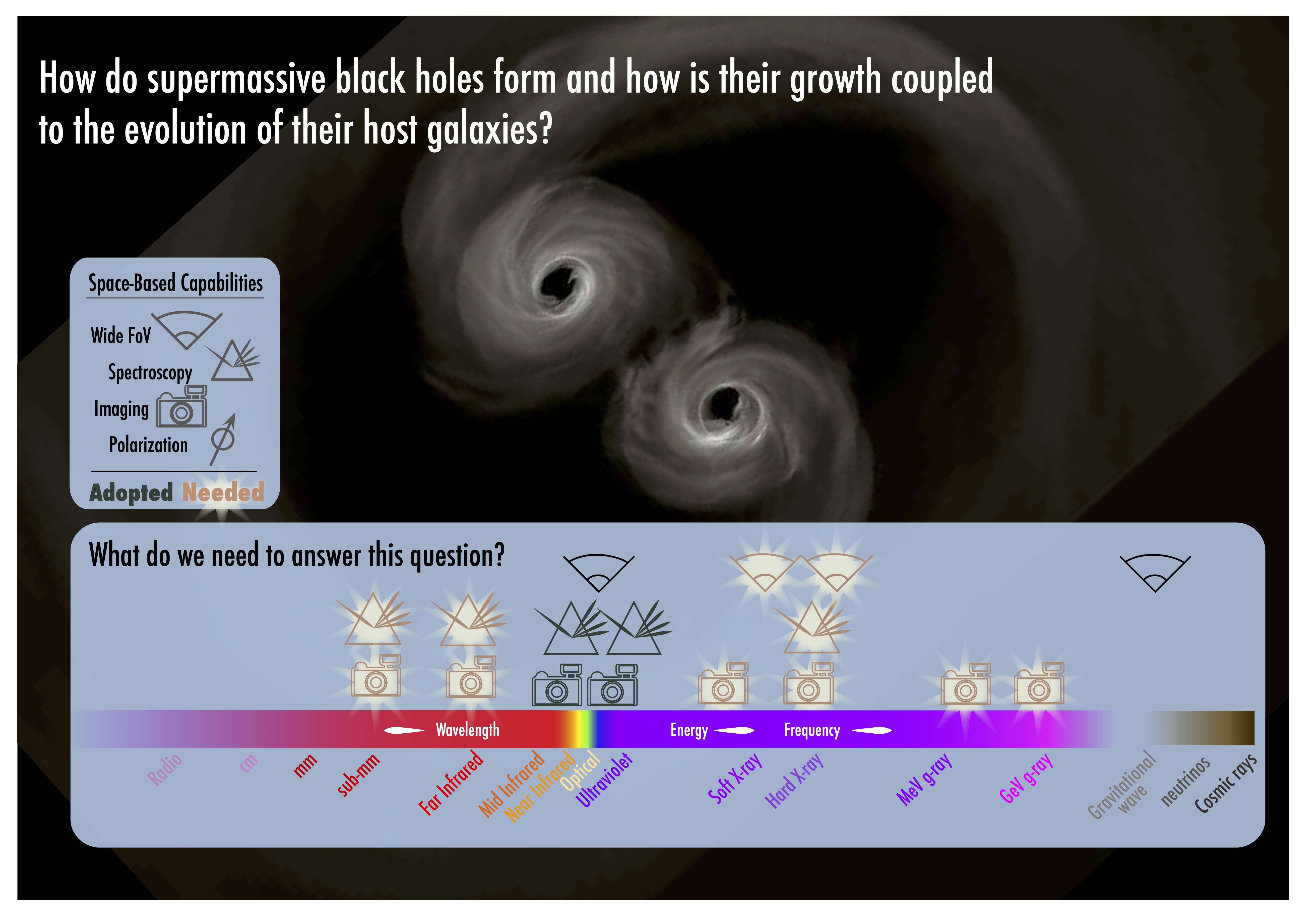}
    \caption{The Astro2020 Decadal science sub-panels outlined the most important scientific questions to be addressed by specific capabilities.  We outline Question D-Q3 from the \textit{New Messengers New Physics} and \textit{Cosmic Ecosystems} themes and the panchromatic capabilities required to answer it. The LISA observatory will provide observations of gravitational waves. The UV, Optical and NIR capabilities are provided by JWST, RST, and HWO, but longer wavelength and high-energy capabilities can be addressed by a coordinated strategic fleet of Flaglet missions.}
    \label{fig:ex2_capabilities}
\end{figure}

Sensitive, sub-arc second resolution, large area gamma-ray and X-ray surveys combined with sub-mm imaging and spectroscopy would provide a large sample of high redshift (z$>$8), active, supermassive block holes and provide new insight into black hole formation models. 
In addition these data would provide detailed information of their host galaxies to understand the onset of their co-evolutionary process. 
UV imaging and spectroscopy would also allow an understanding of BH accretion processes. 
This combination will provide the complete census of supermassive black holes over a broad redshift and mass range, including the obscured sources that are not selectable at optical wavelengths. 
mid- and far-IR infrared spectroscopy are needed for longer-wavelength diagnostic features of obscured growth beyond z$\sim$2.
High-spectral-resolution X-ray kinematic measurements provide a promising way to study feedback mechanism across a broad range of galaxies masses, environments, and evolutionary stage. 
To probe active galaxy winds across high-throughput, high-resolution spectroscopy from the hard X rays through the far ultraviolet (FUV) is needed. 
This diverse combination of capabilities can be achieved by a fleet of synergistic observatories, as shown in Figure~\ref{fig:ex2_capabilities}. 

\paragraph{What are the properties of individual planets, and what processes lead to planetary diversity? (E-Q2):}  Thousands of exoplanets are currently known in the Milky Way, and these other worlds span a wide range of radii, masses, and bulk properties. 
The demographics of this exoplanet sample are beginning to reveal trends that can be linked to general properties like orbital period, host star type, age, evolutionary history, and environment. Deeper context linking the diversity of planets and to planet formation processes requires more information on individual planets at all stages of their formation and evolution. 

\begin{figure}[ht]
    \centering
    \includegraphics[width=0.95\linewidth]{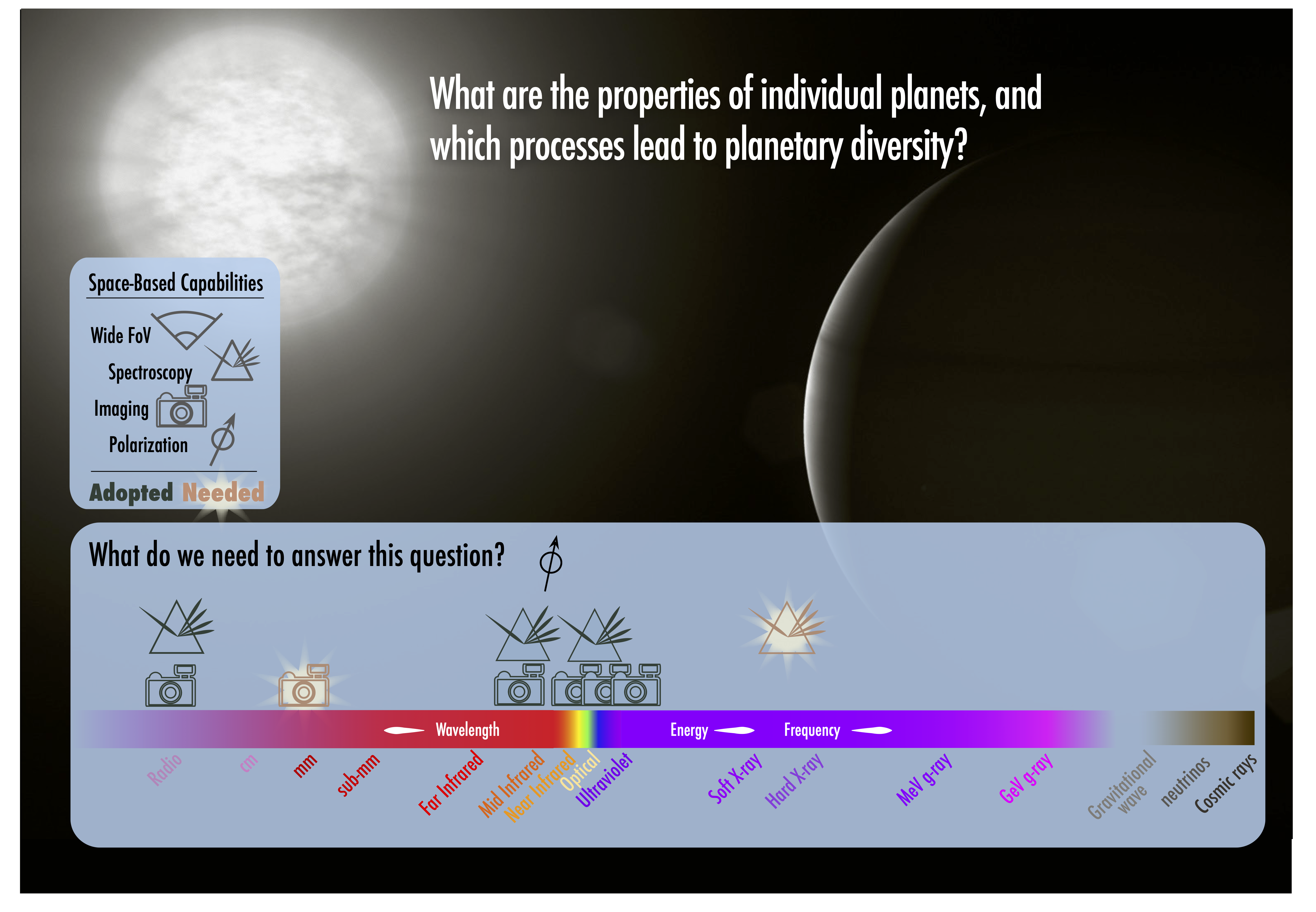}
    \caption{The Astro2020 Decadal science sub-panels outlined the most important scientific questions to be addressed by specific capabilities.  We outline Question E-Q2 from the \textit{Worlds and Suns in Context} theme and the panchromatic capabilities required to answer it. Radio observations can be achieved by ground based telescopes. The UV, Optical and NIR capabilities are provided by JWST, RST, and HWO, but longer wavelength and high-energy capabilities can be addressed by a coordinated strategic fleet of Flaglet missions.}
    \label{fig:ex3_capabilities}
\end{figure}

These measurements require multi-wavelength observations using different observational techniques that leverage multiple observatories. Probing individual planets in the dense, dusty star forming regions while they form requires high resolution imaging and spectroscopy at radio and mm wavelengths that would be achievable with future interferometers. 
Sensitive photometric, spectroscopic, and polarization observations spanning the UV to mid-IR allow access to atmospheric, and in some cases, surface, properties of maturing and mature planets orbiting every kind of star. 
X-ray observations, particularly spectroscopy, provide key contextual information on the energetic activity of exoplanet host stars over their lifetimes, which has a significant impact on their atmosphere evolution. 
Similarly, the study of the impact of short-timescale stellar activity (space weather) on atmosphere characteristics such as radiatively-driven atmospheric loss and photochemistry requires near-simultaneous measurements of both the stellar output and the planet's atmosphere.  
Future characterization of individual planets across the full timescale of their formation and evolution is ideally suited to a panchromatic fleet of observatories with commensurate operations as can be seen in Figure~\ref{fig:ex3_capabilities}.

\paragraph{How do star-forming structures emerge from, and interact with, the diffuse interstellar medium (ISM)? (F-Q1):} The process of star formation spans many orders of magnitude of time and physical scale. Under the influence of gravity and other forces, parsec size interstellar gas clouds collapse into solar system scale cores that ultimately ignite sustained fusion and result in compact stars. These stars then continue to evolve and their energetic emission has significant impact on their local environment. Observing these star formation processes directly is difficult due to the environment that fosters them; large clouds of gas and dust and dense, compact cores both obscure the details and require high-sensitivity and high-resolution remote sensing. 

New breakthroughs in star formation can be achieved through combinations of centimeter wavelength interferometry and far-IR, UV, and X-ray spectroscopy. This wide range of wavelengths and observational regimes is well matched to the series of next generation great observatories described here.

\paragraph{What are the most extreme types of stars and stellar populations? (G-Q1):} 
Stars at the extremes of mass, composition, rotation, and pulsation challenge theoretical models and shape the evolution of galaxies. 
Rare transitional phases, when stars are out of equilibrium, remain poorly observed despite being universal. 
Defining these boundaries is critical for stellar systems, where diverse properties — mass, rotation, magnetic fields, and mass transfer — govern evolution. 
For clusters and stellar populations, the distribution of stellar masses and multiplicity strongly influences their development. 
Updated HR diagrams from Gaia and JWST, combined with theory, are already transforming our understanding of stellar evolution and stellar populations across the Milky Way and local universe.

Advancing this science further requires many parallel capabilities. Highly multiplexed, panchromatic (UV to Mid-IR) spectroscopic surveys will measure stellar temperatures, luminosities, abundances, velocities, and variability at scale, and match the vast photometric and astrometric datasets already in hand.
Panchromatic coverage from X-ray to infrared enables studies of coronal heating, mass loss, chemical abundances, and molecular chemistry, while multi-epoch data reveal binaries, asteroseismic structure, and dynamic atmospheres. Advances in spectropolarimetry will map stellar magnetic fields across the mass spectrum. 
High-resolution imaging and interferometry, reaching 10–100 milliarcsecond to sub-microarcsecond scales, are essential to resolve stars and binaries, measure radii, track stellar motions, and probe thermal and magnetic properties down to solar and substellar scales. A synergistic fleet of observatories with overlapping operational timescales can provide these new capabilities in parallel.

\begin{sidewaystable}[htbp]
\centering
\caption{Astrophysics Panels, Themes, and Questions.}
\label{tab:questions}
\renewcommand{\arraystretch}{1.2}
\begin{tabularx}{\textwidth}{@{}>{\raggedright\arraybackslash}p{3cm} >{\raggedright\arraybackslash}p{2cm} >{\raggedright\arraybackslash}p{5cm} X@{}}
\toprule
\textbf{Panel} & \textbf{Theme} & \textbf{Question} & \textbf{Capability} \\
\midrule
Panel on Compact Objects and Energetic Phenomena & NMNP & 
\textbf{B-Q3:} Why do some compact objects eject material in nearly-light-speed jets, and what is that material made of? & 
OIR imaging \& spectroscopy, X-ray imaging \& polarimetry, hard X-ray imaging \& spectroscopy, MeV polarimetry \& imaging, GeV imaging, TeV neutrinos, cosmic rays \\
\addlinespace

Panel on Cosmology & NMNP & 
\textbf{Q-C2:} What are the properties of dark matter and the dark sector? & 
radio telescopes, large-aperture optical telescopes, high-resolution CMB polarization mapping, GeV telescopes, and TeV-scale Cherenkov telescopes  \\
\addlinespace

Panel on Galaxies & NMNP \& CE & 
\textbf{D-Q3:} How do supermassive black holes form, and how is their growth coupled to the evolution of their host galaxies? & 
Sub-mm imaging \& spectroscopy, far-IR imaging \& spectroscopy, OIR wide field, imaging \& spectroscopy, UV imaging \& spectroscopy, X-ray imaging \& wide field, hard X-ray wide-field \& spectroscopy, MeV imaging, GeV imaging, GW low-frequency \\
\addlinespace

Panel on Exoplanets, Astrobiology and the Solar System & WSC & 
\textbf{E-Q2:} What are the properties of individual planets, and what processes lead to planetary diversity? & 
Radio interferometry \& spectroscopy, mm interferometry, MIR spectroscopy \& photometry, OIR imaging, spectroscopy, polarization, \& photometry, UV imaging, X-ray spectroscopy \\
\addlinespace

Panel on the ISM and Star and Planet Formation & CE & 
\textbf{F-Q1:} How do star-forming structures emerge from, and interact with, the diffuse interstellar medium (ISM)? & 
cm interferometry, FIR spectroscopy, UV spectroscopy, X-ray spectroscopy \\
\addlinespace

Panel on Stars, the Sun and Stellar Populations & WSC \& CE & 
\textbf{G-Q1:} What are the most extreme types of stars and stellar populations? & 
Radio interferometry, MIR spectroscopy, NIR spectroscopy, OIR photometry \& imaging, UV spectroscopy, polarization, \& imaging \\
\bottomrule
\end{tabularx}
\end{sidewaystable}

\subsection{Cross Directorate Synergies}
\paragraph{Planetary Decadal}

The ``Origins, Worlds, and Life: A Decadal Strategy for Planetary Science and Astrobiology 2023-2032", hereafter OWL~\cite{NAP26141}, identifies 12 high-level thematic questions for intensive study across all bodies of the solar system. 
Within these, strategic research activities identified targets, missions, and telescopic observations that could address these thematic questions. Large survey and outer solar system studies are particularly well-suited to space-based telescopes, owing to the high costs and few opportunities for deep space missions combined with the much smaller apparent motions on the sky (while they require moving object capabilities, it is not as demanding as tracking inner solar system objects). 

We grouped the OWL high-level questions into five broad topical areas of interest (see \textbf{Table~\ref{tab:SS}}) approximately corresponding to measurements of (i) protoplanetary disks and interstellar medium (ISM) composition; (ii) outer solar system bodies' dynamics and geology; (iii) outer solar system bodies' composition and aurora; (iv) time domain science; and (v) outer solar system small bodies (Trans-Neptunian Objects (TNOs) and Kuiper Belt Objects (KBOs)). 
We followed the same approach outlined in \textbf{Section~\ref{sec:utility}}. 

\begin{sidewaystable}[htbp]
\centering
\caption{Solar System Questions, Targets, and Capabilities~\cite{NAP26522}.}
\label{tab:SS}
\renewcommand{\arraystretch}{1.2}
\begin{tabularx}{\textwidth}{@{}>{\raggedright\arraybackslash}p{3.5cm} >{\raggedright\arraybackslash}p{7.5cm} >{\raggedright\arraybackslash}p{8cm} X@{}}
\toprule
 \textbf{OWL Question} & \textbf{Topical Area} & \textbf{Capability} \\
\midrule
\textbf{Q1:} Evolution of the Protoplanetary Disk & Protoplanetary disks and ISM composition, Outer solar system bodies' dynamics and geology, Outer solar system bodies' composition and aurora, Time domain science & Very high spectral resolution IR spectra, High resolution OIR imaging, UV and IR spectra, Simultaneous vis through thermal IR imaging\\
\addlinespace
\textbf{Q2:} Accretion in the Outer Solar System & Protoplanetary disks and ISM composition, Outer solar system bodies' dynamics and geology, Outer solar system bodies' composition and aurora, Time domain science, Outer solar system small bodies & Very high spectral resolution IR spectra, High resolution OIR imaging, UV and IR spectra, Simultaneous vis through thermal IR imaging, Wide-field IR imaging\\
\addlinespace
\textbf{Q4:} Impacts and Dynamics & Outer solar system small bodies & Wide-field IR imaging\\
\addlinespace
\textbf{Q7:} Giant Planet Structure and Evolution & Protoplanetary disks and ISM composition, Outer solar system bodies' dynamics and geology, Outer solar system bodies' composition and aurora, Time domain science & very high spectral resolution IR spectra, high resolution OIR imaging, UV and IR spectra, simultaneous vis through thermal IR imaging\\
\addlinespace
\textbf{Q8:} Circumplanetary Systems & Protoplanetary disks and ISM composition, Outer solar system bodies' dynamics and geology, Outer solar system bodies' composition and aurora, Time domain science & Very high spectral resolution IR spectra, High resolution OIR imaging, UV and IR spectra, Simultaneous vis through thermal IR imaging\\
\textbf{Q12:} Exoplanets & Protoplanetary disks and ISM composition, Outer solar system bodies' dynamics and geology, Outer solar system bodies' composition and aurora, Time domain science & Very high spectral resolution IR spectra, High resolution OIR imaging, UV and IR spectra, Simultaneous vis through thermal IR imaging\\

\bottomrule
\end{tabularx}
\end{sidewaystable}

With the currently operating and soon to launch trio of HST, JWST, and RST, many of the telescopic science requirements for three of the five topical areas can already be substantially addressed - namely all three of the ``Outer Solar System" topics, which are already routine targets for HST and JWST. 
For the remaining topics, flatlet-size missions would entirely enable that science to be achieved. 

We note that while a program of targeted planetary missions will likely continue through the Planetary Science Division (missions such as Mars rovers, Dragonfly, Uranus Orbiter and Probe, Enceladus Orbilander etc.), these typically are directed to answer a focused set of science goals for a single object, and in some cases, required in-situ measurements; therefore they do not replace the need for astrophysical space telescopes that can be pointed at multiple bodies in the solar system and research whole categories of objects.

\subsection{Moon-to-Mars, Planetary Defense, and National Security}
The instrument capabilities required to achieve the science goals described in Astro2020 are also highly relevant to emerging NASA priorities. Returning American astronauts to the Moon and then onward to Mars is a core goal of the agency. Successfully and safely achieving this requires both in-situ and remote sensing measurements across a wide range of wavelengths and energies. UV, X-ray, gamma-ray, and particle measurements in low-earth, cis-lunar, and lunar orbit provide deep insight into the space weather environment to characterize, model, and forecast potentially dangerous solar and geomagnetic storms. A suite of panchromatic flatlet-size missions can provide these key measurements while simultaneously making breakthroughs in Decadal science.  

The detection and remote characterization of potentially hazardous near-Earth objects (NEOs) is a core component of planetary defense and is closely linked to exploration of the Earth-Moon-Mars environment. Such measurements improve threat assessments and provide necessary context for hazard mitigation strategies. The multi-wavelength survey measurements needed for NEO detection and high-sensitivity, high-resolution observations needed for NEO characterization are provided by the same instruments on a fleet of next generation Great Observatories that achieve Astro2020 goals. Remote observations of Mars are also a unique contribution achievable by Flaglet-size mission capabilities, particularly multi-wavelength high-resolution imaging and integral field spectroscopy. A synergistic fleet of observatories in low and near-earth orbits can perform routine, whole disk monitoring of the red planet to build a deeper understanding of seasonal changes and Mars weather phenomena which are critical for surface operation safety. In the era of human presence on and around the martian surface, rapid situational awareness assessments provided by such a fleet can complement such activity and ensure astronaut health and safety.

The instruments flown on a suite of Flaglet-size missions will also contribute directly to national security. Multiple instruments covering a wide range of energies and wavelengths provide space situational awareness and could be used to detect and assess anomalous activity and threats. In addition, the technologies developed to achieve these missions will have wide ranging benefit through infusion into other government agencies and industry and through direct partnerships. This includes advanced detector technologies, low-noise and noiseless electronics, quantum sensing devices, next generation communications and ground systems, large-scale data archives, and computing infrastructure and analysis algorithms that leverage AI and machine learning. 

\section{A Flaglet Great Observatory Program}

Strategic program management enables improved science return through efficient application of money, facilities, the NASA workforce, and the larger US science community. 
The following subsections explain how these factors are inseparable from a successful program. 

Based on our analysis, from four to six Flaglet-size missions will enable NASA to address $\sim$80\% of the Astro2020 science gaps left after JWST, RST, and HWO. 
Strategic programmatic design choices would lead to combined measurement capabilities, which we expect to reduce the envisaged fleet to four such missions. 
Below we describe a program comprising four flatlet-size missions. 

\subsection{Budget Profile}

Mission budgets are far from constant with time, and generally follow a well documented~\cite{NAPLargeMission} profile ramping up through formulation to a peak during Phase C and then dropping towards launch and remaining at a fairly constant level during operations. This time evolution means that while a constant budget line is a poor match for a single mission, a properly sequenced series of similar cost missions can fit well. 

\begin{figure}[ht!]
    \centering
    \includegraphics[width=0.95\linewidth]{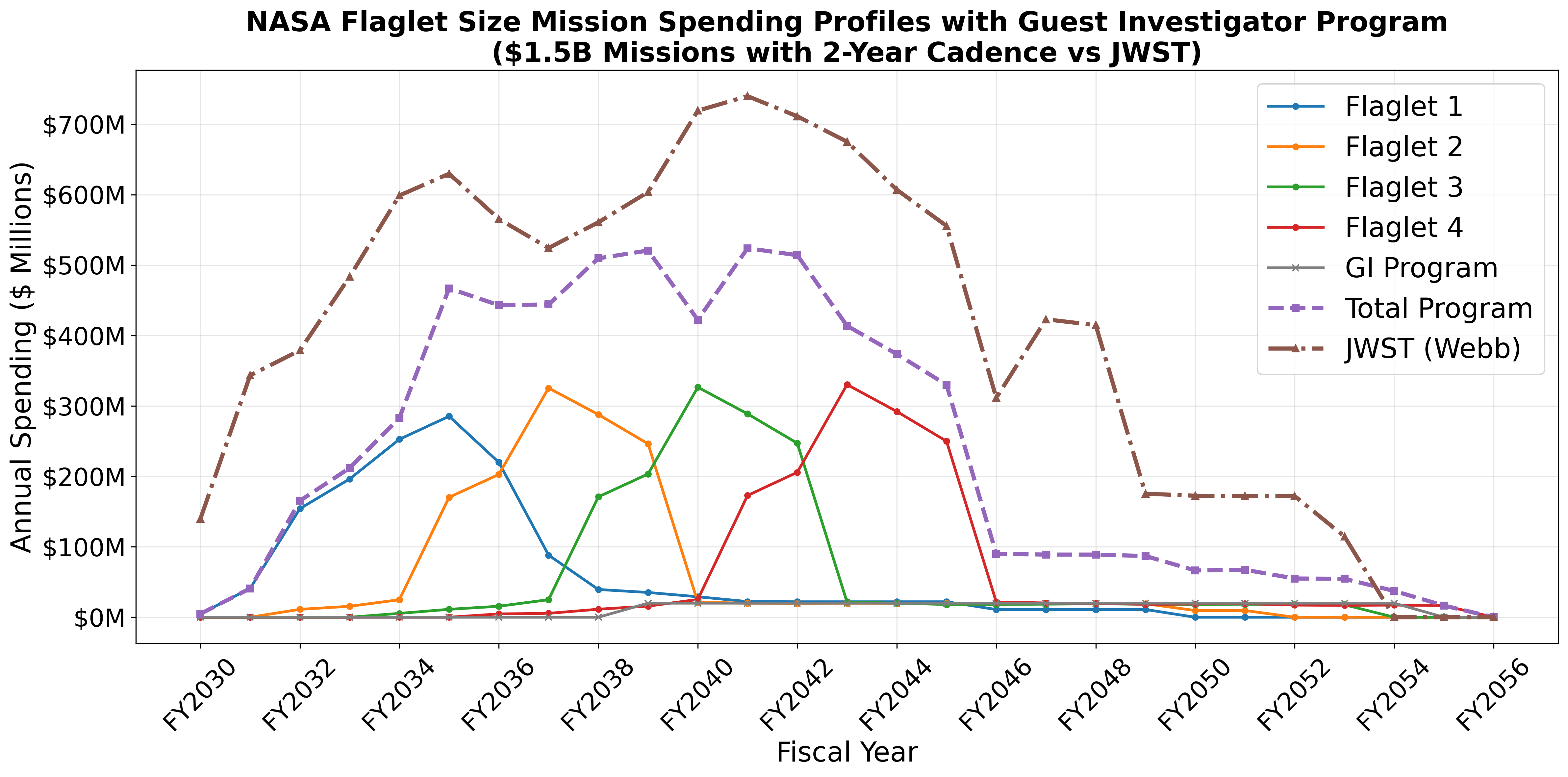}
    \caption{The budget profile of our proposed fleet of Flaglet-size missions demonstrates that four strategic missions with staggered starts and overlapping operations deliver both transformative science and a feasible spending profile. The cost includes four Flaglets ramping up successively every two years, each at a cost cap of \$1.5B. We also include a joint \$20M per year GI/GO program, that would continue throughout the lifetime of the missions. The complementary nature of these missions creates a robust observational framework and a strategic foundation for future astrophysics. For illustration, we include the annual program cost of JWST adjusted to 2020 dollars~\cite{jwst}. }
    \label{fig:2yearstart}
\end{figure}

\begin{figure}[ht!]
    \centering
    \includegraphics[width=0.95\linewidth]{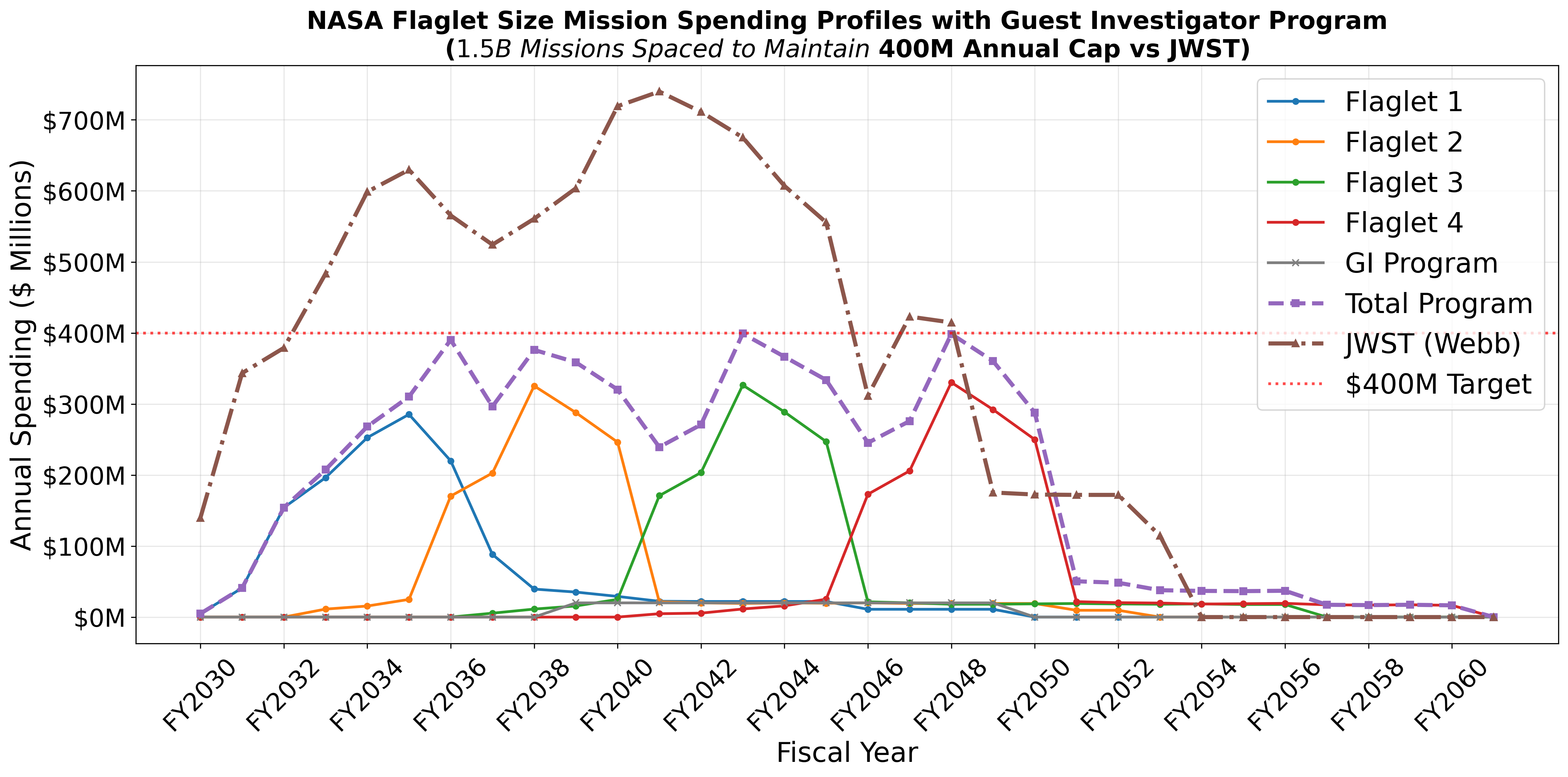}
    \caption{The budget profile of our proposed fleet of Flaglet-size missions demonstrates that four strategic missions with staggered starts and overlapping operations deliver both transformative science and a feasible spending profile. The cost includes four flaglets ramping up successively on a cadence that caps the program cost at \$400M, with each mission having a cost cap of \$1.5B. We also include a joint \$20M per year GI/GO program, that would continue throughout the lifetime of the missions. The complementary nature of these missions creates a robust observational framework and a strategic foundation for future astrophysics. For illustration, we include the annual program cost of JWST adjusted to 2020 dollars~\cite{jwst}. }
    \label{fig:400Mcap}
\end{figure}

We took two approaches to calculating the programmatic budget profile. One approach maintains a 2-year cadence between new mission starts (\textbf{Figure \ref{fig:2yearstart}, Table \ref{tab:mission_summary_2yr}}), and the other maintains a \$400M cost cap annually for the program (\textbf{Figure \ref{fig:400Mcap}, Table \ref{tab:mission_summary}}).
In both cases, we assumed a total mission cost cap of \$1.5B, with no adjustment for inflation, and a \$20M annual General Investigator/Observer (GI/GO) program for at least 10 years post launch. 
For illustration, we include the annual program cost of JWST adjusted to 2020 dollars~\cite{jwst}. 
The project started in 2003, and that start date was shifted to FY2030 to align with the fleet proposed.  
We took the Space Infrared Interferometric Telescope (SPIRIT) mission concept~\cite{spirit} as a template for the cost profile of an individual mission, assumed mature technologies and a compressed 1-year pre-formulation period, scaled the total mission cost to \$1.5B, and stretched the planned 5-year operational period to 20 years. 
The cost for launch services was estimated to be \$400M in 2010 and would likely be lower today - especially given advances outlined in Section~\ref{sec:capabilities}. 
Subsequent mission cost profiles allow for up-front technology maturation investments.

\begin{table}[ht!]
\centering
\caption{Flaglet Mission Program Summary (2-Year Staggered Start). The missions would be planned to support a lifespan of 20 years, a \$1.5B cost cap, and maintain an at least~\$20M per year joint GI/GO program throughout their lifetime. }
\label{tab:mission_summary_2yr}
\begin{tabular}{|l|c|c|c|}
\hline
\multicolumn{4}{|c|}{\textbf{Individual Mission Details}} \\
\hline
\textbf{Mission} & \textbf{Start Year} & \textbf{Peak Spending} & \textbf{Peak Year}  \\
\hline
Flaglet 1 & FY2030 &  \$285.4M & FY2035  \\
Flaglet 2 & FY2032 &  \$325.5M & FY2037  \\
Flaglet 3 & FY2034 &  \$326.5M & FY2040  \\
Flaglet 4 & FY2036 &  \$330.3M & FY2043  \\
\hline
\multicolumn{4}{|c|}{\textbf{Total Future Great Observatory Program Summary}} \\
\hline
\multicolumn{2}{|l|}{Program Duration} & \multicolumn{2}{c|}{FY2030 -- FY2056 (27 years)} \\
\multicolumn{2}{|l|}{Mission Duration} & \multicolumn{2}{c|}{20 years ($>$10 year Phase E) } \\
\multicolumn{2}{|l|}{Total Program Cost} & \multicolumn{2}{c|}{\$6320M (\$320 for GI)} \\
\multicolumn{2}{|l|}{Combined Peak Spending} & \multicolumn{2}{c|}{\$523.9M in FY2041} \\
\multicolumn{2}{|l|}{Mission Cadence} & \multicolumn{2}{c|}{2-year staggered start} \\
\hline
\end{tabular}
\end{table}

\begin{table}[ht!]
\centering
\caption{Flaglet Mission Program Summary (\$400M annual cap). The missions would be planned to support a lifespan of 20 years, a \$1.5B cost cap, and maintain an at least~\$20M per year joint GI/GO program throughout their lifetime.}
\label{tab:mission_summary}
\begin{tabular}{|l|c|c|c|}
\hline
\multicolumn{4}{|c|}{\textbf{Individual Mission Details}} \\
\hline
\textbf{Mission} & \textbf{Start Year} & \textbf{Peak Spending} & \textbf{Spacing} \\
\hline
Flaglet 1 & FY2030 & \$285.4M & -- \\
Flaglet 2 & FY2033 & \$325.5M & +3 years \\
Flaglet 3 & FY2037 & \$326.5M & +4 years \\
Flaglet 4 & FY2041 & \$330.3M & +4 years \\
\hline
\multicolumn{4}{|c|}{\textbf{Program Summary}} \\
\hline
\multicolumn{2}{|l|}{Program Duration} & \multicolumn{2}{c|}{FY2030 -- FY2061 (32 years)} \\
\multicolumn{2}{|l|}{Mission Duration} & \multicolumn{2}{c|}{20 years ($>$10 year Phase E) } \\
\multicolumn{2}{|l|}{Total Program Cost} & \multicolumn{2}{c|}{\$6,220M (\$220 GI)} \\
\multicolumn{2}{|l|}{Peak Combined Spending} & \multicolumn{2}{c|}{\$398.1M (under \$400M target)} \\
\hline
\end{tabular}
\end{table}

\subsection{Facilities and Workforce}

NASA, its Centers, and its partners have the capacity to build multiple large missions in parallel.  Each mission has different personnel, infrastructure, and financial requirements primarily as a function of mission phase and secondarily as a function of the previously mentioned set of capabilities. In the preceding section, roughly one mission is in each phase (A-D) at any given time, which avoids institutional and budgetary oversubscription as well as prevents the unavoidable atrophy of capabilities and facilities driven by long gaps between use. 

\subsection{Joint General Observer/Investigator Program}

This program also includes a joint General Observer/Investigator program of~\$20M per year starting 5 years after the first mission enters the implementation phase. 
We assume a constant~\$20M until the end of the program and that it is shared among observatories. 
This line is independent of operational costs for the observatory and is meant to go to the community. 

\subsection{Multi-Agency and International Coordination}
The key science questions we seek to answer are agnostic to the techniques required and span all wavelengths and messengers, including those best studied via ground-based facilities. 
For example, neutrinos with energies below the $10^{18}$ eV range are best studied from the ground (or deep below), as they require instrumenting large volumes of water or ice. 
Similarly, gamma rays with energies above some 10's GeV are best studied from the ground because of the collecting areas required. 
More practically, in many electromagnetic bands where observations are feasible in principle from either space or ground, ground-based observatories offer advantages in the cost of deployment and accessibility, and in the size scales that can be reached.  

Planning for a fleet of Flaglet-size observatories will be most effective if conducted as a strategic exercise including both other U.S. funding agencies (particularly NSF and DOE) and international partners (e.g., ESA, JAXA, CSA). 
Just as the exercise above includes the future NASA Flagships that are part of the program of record, evaluating the science potential of Flaglet-size mission concepts must be done in the context of other current and planned facilities. 
Strategic planning between NASA and other funding agencies is required to create coordination of timelines for facilities whose simultaneous operation is needed to optimize science return. 
While it has proven challenging in the past to coordinate even between different agencies within the U.S., the size of investments being made and the multi-year or longer timescales for deploying new facilities dictates that coordinated strategic planning between agencies and between countries is required to maximize science return. 
As one example, uncertainty in the operations timelines for the current ground-based gravitational-wave observatories has made it difficult for proposers to NASA SMEX and MIDEX AOs to argue for complementary space-based observatories. 
The Astronomy and Astrophysics Advisory Committee (AAAC) may be well positioned to play a role in helping the U.S. agencies build the necessary coordination. 

A series of Flaglet-scale Great Observatories would provide numerous opportunities to continue and build global partnerships via international contributions to the missions. These contributions may take the form of instruments that provide unique or enhanced capabilities. 
They may include ground stations that enable rapid commanding and handling of increased data volumes. 
Lastly, they can include data centers that distribute the workload; while future missions will likely rely on cloud-based computing, there is still a need to locate the data in geographical proximity to the scientists who want to work with it.

\section{Summary of Findings}

Much of the decadal survey science needs a panchromatic fleet and a broad program of
“... missions of all scales, national and international, designed to view the universe in a multiplicity of complementary ways are now essential to progress in modern astrophysics.” \footnote{p 1-12; Section 2; p 5-1; p1-9}. 
Therefore, we see the need for a strategic program to address the decadal science because it cannot be achieved with a single mission. 
The scientific scope of the decadal questions drive the need for contemporaneous operations of the missions in a program.

The findings presented in this white paper demonstrate that a strategic program of Flaglet-size missions represents the optimal approach to addressing the comprehensive scientific questions outlined in the Astro2020 decadal survey. 
Our analysis shows that with approximately four strategically designed and curated missions complementing current and planned Flagship observatories, the astrophysics community can address a significant fraction of the priority scientific questions identified by the decadal panel.

The proposed Flaglet mission program bridges the critical gap between smaller Explorers and large Flagship missions, offering the ideal balance between scientific capability, cost-effectiveness, and implementation timeline. 
By establishing a coordinated fleet approach rather than pursuing missions in isolation, NASA can maximize the scientific return on investment through complementary capabilities, contemporaneous observations, and shared infrastructure.

This strategic approach provides multiple benefits beyond pure scientific output:
\begin{itemize}
    \item Provides stability and predictability for technology development
    \item Leverages advances in commercial space capabilities and cross-directorate synergies
    \item Enhances U.S. leadership in space-based astrophysics
    \item Establishes a sustainable cadence of missions aligned with career timelines and opportunities for workforce development 
\end{itemize}

Moving forward, the coordinated Flaglet fleet architecture we envision is a natural and timely fit for NASA's ASTRA initiative. 
ASTRA is explicitly designed to advance mission concept studies, mature enabling technologies, and maintain a pipeline of strategic mission concepts ahead of formal project formulation, which is precisely the groundwork needed to realize a coherent, multi-mission fleet strategy. 
We suggest a community-led mission study under ASTRA to define the Flaglet fleet architecture and expand upon the analyses peformed here: characterizing mission configurations, identifying capability and wavelength-regime trades, and establishing the operational synergies that maximize science return in concert with Flagship missions. 
With ASTRA community engagement activities anticipated to begin as early as mid-2026, the timing is ideal to bring the astrophysics community together around this vision and ensure that the coordinated fleet approach is positioned as a central pillar of NASA's next-generation strategic planning. 
Through coordination and thoughtful program management, we can maximize the scientific return on NASA's astrophysics investments for decades to come.

The authors gratefully acknowledge the contributions of Brendan Crill, Thomas Essinger-Hileman, and Shaul Hanany to this white paper on CMB science and capabilities. 

\newpage
\bibliography{main}   

@BOOK{NAP26522,
  author    = "{National Academies of Sciences Engineering and Medicine}", 
  title     = "Origins, Worlds, and Life: A Decadal Strategy for Planetary Science and Astrobiology 2023-2032",
  isbn      = "978-0-309-47578-5",
  doi       = "10.17226/26522",
  url       = "https://nap.nationalacademies.org/catalog/26522/origins-worlds-and-life-a-decadal-strategy-for-planetary-science",
  year      = 2023,
  publisher = "The National Academies Press",
  address   = "Washington, DC"
}

@BOOK{NAP26141,
  author    = "{National Academies of Sciences Engineering and Medicine}",
  title     = "Pathways to Discovery in Astronomy and Astrophysics for the 2020s",
  isbn      = "978-0-309-46734-6",
  doi       = "10.17226/26141",
  url       = "https://nap.nationalacademies.org/catalog/26141/pathways-to-discovery-in-astronomy-and-astrophysics-for-the-2020s",
  year      = 2023,
  publisher = "The National Academies Press",
  address   = "Washington, DC"
}

@techreport{harwit1986,
  author = {Harwit, M. and Neal, V.},
  title = {The great observatories for space astrophysics},
  institution = {NASA},
  year = {1986},
  number = {NTRS 19860015241},
  url = {https://ntrs.nasa.gov/citations/19860015241}
}

@misc{planetarysociety,
  title = "{Historical NASA Budget Data}",
  howpublished = {\url{https://docs.google.com/spreadsheets/d/e/2PACX-1vTU9FhDV4U6X4suHtvoiMLYDN-y56ipoGh-N7n9fNq7BW1PiMsx5fVlj10LsgvTYVbu3CiUDO_WD0We/pubhtml}},
  author = "{The Planetary Society}",
  year = 2025, 
  note = {Accessed: 2025-09-17}
}

@misc{spirit,
  title = "{The Space Infrared Interferometric Telescope (SPIRIT):A Far-IR Observatory for High-resolution Imaging and Spectroscopy}",
  howpublished = {\url{https://asd.gsfc.nasa.gov/spice/documents/SPIRIT_PPP_RFI_final.pdf}},
  author = "{D. Leisawitz}",
  year = 2009, 
  note = {Accessed: 2026-05-01}
}

@misc{jwst,
  title = "{How much does the James Webb Space Telescope cost?}",
  howpublished = {\url{https://www.planetary.org/articles/cost-of-the-jwst}},
  author = "{The Planetary Society}",
  year = 2021, 
  note = {Accessed: 2025-09-17}
}

@BOOK{NAPLargeMission,
  author    = "{National Academies of Sciences Engineering and Medicine}", 
  title     = "Powering Science: NASA's Large Strategic Science Missions",
  isbn      = "978-0-309-46386-7",
  doi       = "10.17226/24857",
  url       = "https://nap.nationalacademies.org/catalog/24857/powering-science-nasas-large-strategic-science-missions",
  year      = 2017,
  publisher = "The National Academies Press",
  address   = "Washington, DC"
}

@article{suborbitals,
author = {Drew M. Miles},
title = {{Great observatories maturation: a review of NASA astrophysics development through suborbital rocket and balloon programs}},
volume = {11},
journal = {Journal of Astronomical Telescopes, Instruments, and Systems},
number = {4},
publisher = {SPIE},
pages = {042220},
year = {2025},
doi = {10.1117/1.JATIS.11.4.042220},
URL = {https://doi.org/10.1117/1.JATIS.11.4.042220}
}

@misc{flaglets,
  title = "{ASSESSING SCIENTIFIC PRODUCTIVITY OF NASA’S SPACE MISSIONS BY COST}",
  howpublished = {\url{https://www.hou.usra.edu/meetings/lpsc2026/pdf/1606.pdf}},
  author = "{The Planetary Society}",
  year = 2026, 
  note = {57th LPSC (2026) Accessed: 2026-04-28}
}

@misc{fleet,
  title = "{Astrophysics Fleet Chart}",
  howpublished = {\url{https://science.nasa.gov/astrophysics/astrophysics-fleet-chart/}},
  author = "{NASA}",
  year = 2025, 
  note = {Updated: June 2025; Accessed: 2026-04-28}
}
\bibliographystyle{abbrvnat}   

\end{document}